\documentclass{mem}
\usepackage{natbib}\usepackage{txfonts}\usepackage{balance}
\usepackage{graphicx}
\usepackage[a4paper]{hyperref}
\idline{76}{282}
\begin{document}
\def\teff{$T\rm_{eff }$}
\def\kms{$\mathrm {km s}^{-1}$}
\newcommand{\mD}{\ensuremath{\left\langle\mathrm{3D}\right\rangle}}

\title{3D hydrodynamical simulations of stellar photospheres with the {\tt CO$^5$BOLD} code}
   \subtitle{Photometric colors of a late-type giant}

\author{
A. \,Ku\v{c}inskas\inst{1,2},
H.-G. \,Ludwig\inst{3},
E. \,Caffau\inst{3},
\and M. \,Steffen\inst{4}
          }

   \offprints{A. Ku\v{c}inskas}

\institute{
Institute of Theoretical Physics and Astronomy, Go\v{s}tauto 12, Vilnius LT-01108, Lithuania \email{ak@itpa.lt}
\and
Vilnius University Astronomical Observatory, \v{C}iurlionio 29, Vilnius LT-03100, Lithuania
\and
GEPI - CIFIST, Observatoire de Paris-Meudon, 5 place Jules Janssen, 92195 Meudon Cedex , France
\and
Astrophysikalisches Institut Potsdam, An der Sternwarte 16, D-14482 Potsdam, Germany
}

\authorrunning{A. Ku\v{c}inskas}

\titlerunning{3D photometric colors of a late-type giant}

\abstract{
We present synthetic broad-band photometric colors of a late-type giant located close to the RGB tip ($T_{\rm eff}\approx3640$\,K, $\log g=1.0$ and ${\rm [M/H]}=0.0$). Johnson-Cousins-Glass \emph{BVRIJHK} colors were obtained from the spectral energy distributions calculated using 3D hydrodynamical and 1D classical stellar atmosphere models. The differences between photometric magnitudes and colors predicted by the two types of models are significant, especially at optical wavelengths where they may reach, e.g., $\Delta V\approx0.16$, $\Delta R\approx0.13$ and $\Delta (V-I)\approx0.14$, $\Delta (V-K)\approx0.20$. Differences in the near-infrared are smaller but still non-negligible (e.g., $\Delta K\approx 0.04$). Such discrepancies may lead to noticeably different photometric parameters when these are inferred from photometry (e.g., effective temperature will change by $\Delta T_{\rm eff}\approx60$\,K due to difference of $\Delta (V-K)\approx0.20$).


}

\maketitle{}

\begin{figure*}[t!]
\begin{centering}
\resizebox{12.5cm}{!}{\includegraphics[clip=true]{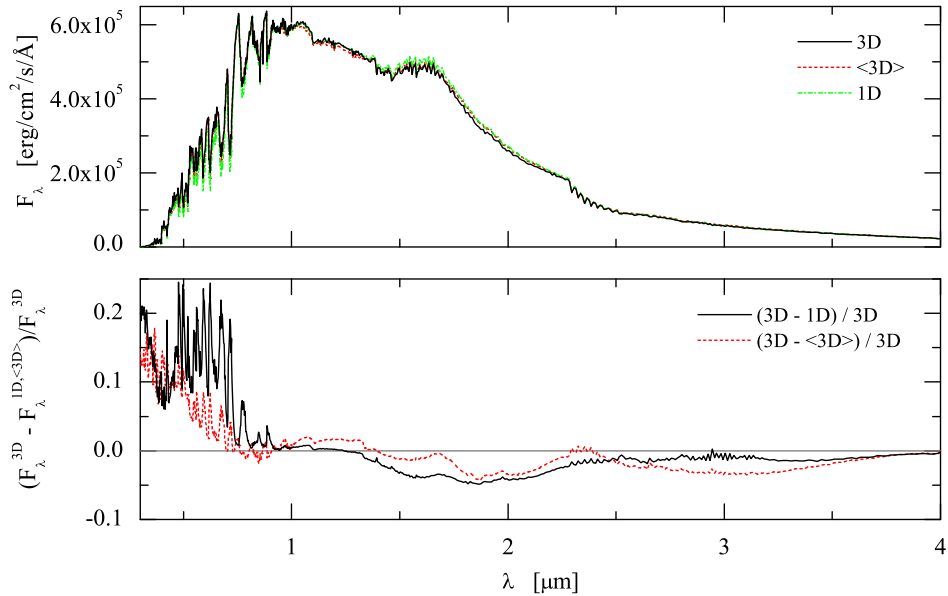}}
\caption{\footnotesize
Top: spectral energy distributions corresponding to the 3D (black solid line), ${\rm <3D>}$ (3D global average, gray/red dashed line) and 1D (gray/green dot-dashed line) models of a late-type giant ($T_{\rm eff}\approx3640$\,K, $\log g=1.0$ and ${\rm [M/H]}=0.0$). Bottom: relative differences between the spectral energy distributions: $({\rm 3D}-{\rm 1D})/{\rm 3D}$ (black solid line) and $({\rm 3D}-\mD)/{\rm 3D}$ (gray/red dashed line).}
\label{SEDs}
\end{centering}
\end{figure*}

\section{Introduction}

Late-type giants are important tracers of intermediate age and old stellar populations in the Galaxy and beyond. Thus, the availability of reliable stellar atmosphere models is of utmost importance for understanding the structure and evolution of these stars and their host populations. The advent of the 3D hydrodynamical codes allowed to accommodate a more realistic treatment of non-stationary phenomena (e.g., convection) in the stellar atmosphere modeling and thus it is natural to expect that new 3D hydrodynamical models may provide important insights about observable properties  and the interior structures of late-type giants. Indeed, recent work of \citet{CAT07}, \citet{C08} demonstrates that there are significant differences in the spectral line strengths predicted by the 3D hydrodynamical and 1D classical stellar atmosphere models. This leads to substantial discrepancies between the elemental abundances of various chemical species derived with the classical 1D and 3D hydrodynamical stellar atmosphere models \citep[see][for details]{CAT07}.

Still, the extent of differences in the global spectral properties predicted by the 1D classical and 3D hydrodynamical models is largely unknown. In one of our previous studies we have used a simplified approach to show that significant differences can be expected in the photometric colors predicted by the two types of models \citep{KHL05}. Whether these conclusions would also hold with the photometric colors based on the results of full 3D spectral synthesis still had to be confirmed.

In this study we extend the previous work and calculate synthetic photometric colors of a late-type giant from the full 3D spectral energy distributions. For this purpose we use a model of a late-type giant which is significantly cooler than those considered by Collet et al. (2007), with its atmospheric properties typical to those of solar-metallicity late-type giants located close to the RGB tip.

\begin{figure*}[t!]
\begin{centering}
\resizebox{10cm}{!}{\includegraphics[clip=true]{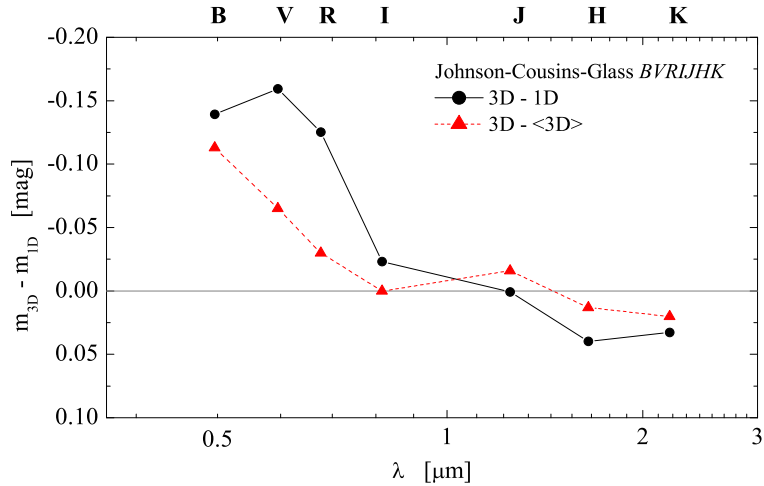}}
\caption{\footnotesize
Differences between the photometric magnitudes of a late-type giant predicted by 3D hydrodynamical and 1D classical models: ${\rm 3D}-{\rm 1D}$ (black solid line) and ${\rm 3D}-\mD$ (red/gray dashed line).}
\label{colors-BVRIJHK}
\end{centering}
\end{figure*}

\section{The models: 3D hydrodynamical and 1D classical}

The 3D hydrodynamical model of a late-type giant was calculated with the {\tt CO$^5$BOLD} stellar atmosphere code \citep{FSD02,FSWL03,WFSLH04}. Atmospheric parameters of the model are: $T_{\rm eff}\approx3640$\,K, $\log g=1.0$ and ${\rm [M/H]}=0.0$. The model utilizes a Cartesian grid of $150\times150\times151$ points ($x,y,z$, respectively) which corresponds to the spatial dimensions of $15.6\times15.6\times8.6$\,Gm or $\sim22.6\times22.6\times12.4$ $R_{\odot}$. The simulation run was performed assuming solar chemical composition. Continuum opacities were taken from the {\tt MARCS} stellar atmosphere package \citep{GEK08} and were grouped into 5 opacity bins according to the procedure described by \citet{N82}, \citet{LJS94} and \citet{VBS04}.

A comparison 1D classical model was calculated using the {\tt LHD} stellar atmosphere code, employing the same atmospheric parameters as those used with the {\tt CO$^5$BOLD} model. The equation of state and opacities used with the {\tt LHD} and {\tt CO$^5$BOLD} models were identical.

\section{Synthetic photometric colors}

The {\tt CO$^5$BOLD} simulation run produces time-series of individual 3D model atmospheres, so-called 'snapshots'. In order to minimize computational costs we selected 15 snapshots (equidistantly spaced in time) that are representative of the statistical properties of a full snapshot ensemble.

To calculate photometric colors of a late-type giant we utilized information about the line-blanketed emergent radiative flux, $F_{\nu}(x,y)$, which was calculated at each horizontal grid point of a given 3D model snapshot using the code {\tt NLTE3D} (Steffen et al. 2009, in prep.). Continuous opacities used in the {\tt NLTE3D} calculations are taken from the {\tt Linfor3D} spectrum synthesis code \citep{SLW08}, and are based on the same chemical abundances as the opacities used with the {\tt CO$^5$BOLD} simulations, while line blanketing was taken into account by implementing opacity distribution functions from \citet[]['little division' ODFs]{CK03}. The obtained spectral energy distributions sample the wavelength interval 300-4000\,nm at 672 wavelength points.

The opacities applied in the calculation of energy distributions are not fully consistent with the opacities used in the 3D model run. This leads to a mismatch between the total emergent flux of the spectral energy distributions and that of underlying 3D model. We therefore corrected the total flux of 3D, \mD\, 1D spectral energy distributions to a nominal value of the effective temperature of 3640\,K by scaling each distribution by a wavelength-independent factor.

The broad-band Johnson-Cousins-Glass colors were calculated from the spectral energy distributions using filter definitions from \citet{B90} for the Johnson-Cousins \emph{BVRI} bands and \citet{BB88} for the Johnson-Glass \emph{JHK} bands. Instrumental magnitudes were converted to the standard Johnson-Cousins-Glass system using zero points derived from the synthetic colors of Vega according to a procedure described in \citet{KHL05}.

\section{Results and discussion}

It is evident that, compared to the 1D model, the 3D hydrodynamical model produces more flux in the blue part of the spectrum and less in the red/near-infrared (Fig. \ref{SEDs}). This leads to noticeable flux differences in the photometric bands, e.g., $\Delta V\approx0.16$, $\Delta R\approx0.13$ (Fig.\ref{colors-BVRIJHK}). Color differences are significant too, e.g., $V-I\approx0.14$, $V-K\approx0.20$.

The discrepancies between the predictions of 3D and 1D models are considerably larger than typical photometric errors and may cause noticeable differences in the photospheric parameters derived using photometric colors. For instance, a difference in color of $\Delta (V-K)\approx0.20$ would change the inferred effective temperature by $\Delta T_{\rm eff}\approx60$\,K. This is comparable to the typical error in the photometrically derived effective temperatures used in stellar abundance work and thus may, in turn, introduce an error in the derived elemental abundances of up to $\sim0.1$\,dex.

Evidently, horizontal inhomogeneities play an important role in defining the shape of 3D spectral energy distributions, as reflected by the differences between the spectral energy distributions and photometric colors corresponding to the full 3D and horizontally averaged 3D models, 3D--\mD\ (Fig. \ref{SEDs}, \ref{colors-BVRIJHK}). Their influence becomes increasingly more important at shorter wavelengths, due to higher sensitivity of the source function to the temperature fluctuations.

\section{Conclusions}

The results obtained allow us to conclude that convection indeed plays an important role in defining the intrinsic atmospheric structures and observed properties of late-type giants. Specifically, we find that spectral energy distributions and photometric colors of a late-type giant produced with 3D hydrodynamical and 1D classical stellar atmosphere models are substantially different. Differences in photometric magnitudes and colors are considerably larger than typical photometric errors (e.g., $\Delta V\approx0.16$, $\Delta (V-K)\approx0.20$). These differences may result in further discrepancies, for instance, in the photospheric parameters derived from photometric colors (e.g., a difference of $\Delta (V-K)\approx0.20$ will change the estimated effective temperature by $\Delta T_{\rm eff}\approx60$\,K). Obviously, this may have direct consequences to any photometric work that relates to late-type giants and thus once again stresses the importance of 3D hydrodynamical model atmospheres to be used in the interpretation of observational data.

\begin{acknowledgements}
AK is grateful to the IAU for a travel grant which helped to attend the event. This work was supported in part by grants from the Lithuanian Science and Studies Foundation (V-19/2009) and the bilateral Lithuanian-Ukrainian programme (31V-153). EC and HGL acknowledge support from EU contract MEXT-CT-2004-014265 (CIFIST).
\end{acknowledgements}


\bibliographystyle{aa}

\end{document}